\documentclass[final]{cvpr}

\usepackage{times}
\usepackage{epsfig}
\usepackage{graphicx}
\usepackage{amsmath}
\usepackage{amssymb}
\usepackage{dblfloatfix}
\usepackage[table,x11names]{xcolor}

\usepackage[pagebackref=true,breaklinks=true,colorlinks,bookmarks=false]{hyperref}

\def\cvprPaperID{****} 
\def\confYear{CVPR 2021}
\definecolor{grayhighlight}{RGB}{213,229,255}

\newsavebox\CBox
\def\textBF#1{\sbox\CBox{#1}\resizebox{\wd\CBox}{\ht\CBox}{\textbf{#1}}}

\begin{document}

\title{
\vspace{-8mm}
Fast and Accurate Quantized Camera Scene Detection on Smartphones,\\ Mobile AI 2021 Challenge: Report}
\author{
Andrey Ignatov \and Grigory Malivenko \and Radu Timofte \and
Sheng Chen \and Xin Xia \and Zhaoyan Liu \and Yuwei Zhang \and Feng Zhu \and
Jiashi Li \and Xuefeng Xiao \and Yuan Tian \and Xinglong Wu \and
Christos Kyrkou \and
Yixin Chen \and Zexin Zhang \and Yunbo Peng \and Yue Lin \and
Saikat Dutta \and Sourya Dipta Das \and Nisarg A. Shah \and Himanshu Kumar \and
Chao Ge \and Pei-Lin Wu \and Jin-Hua Du \and
Andrew Batutin \and Juan Pablo Federico \and Konrad Lyda \and Levon Khojoyan \and
Abhishek Thanki \and Sayak Paul \and
Shahid Siddiqui
}

\maketitle

\maketitle

\begin{abstract}

Camera scene detection is among the most popular computer vision problem on smartphones. While many custom solutions were developed for this task by phone vendors, none of the designed models were available publicly up until now. To address this problem, we introduce the first Mobile AI challenge, where the target is to develop quantized deep learning-based camera scene classification solutions that can demonstrate a real-time performance on smartphones and IoT platforms. For this, the participants were provided with a large-scale CamSDD dataset consisting of more than 11K images belonging to the 30 most important scene categories. The runtime of all models was evaluated on the popular Apple Bionic A11 platform that can be found in many iOS devices. The proposed solutions are fully compatible with all major mobile AI accelerators and can demonstrate more than 100-200 FPS on the majority of recent smartphone platforms while achieving a top-3 accuracy of more than 98\%. A detailed description of all models developed in the challenge is provided in this paper.

\end{abstract}
{\let\thefootnote\relax\footnotetext{%
\hspace{-5mm}$^*$
Andrey Ignatov, Grigory Malivenko and Radu Timofte are the Mobile AI 2021 challenge organizers \textit{(andrey@vision.ee.ethz.ch, grigory.malivenko @gmail.com, radu.timofte@vision.ee.ethz.ch)}. The other authors participated in the challenge. Appendix \ref{sec:apd:team} contains the authors' team names and affiliations. \vspace{2mm} \\ Mobile AI 2021 Workshop website: \\ \url{https://ai-benchmark.com/workshops/mai/2021/}
}}

\vspace{-2mm}
\section{Introduction}

\begin{figure*}[t!]
\resizebox{\linewidth}{!}
{
\large
\begin{tabular}{cccccc}
   \includegraphics[width=0.32\linewidth]{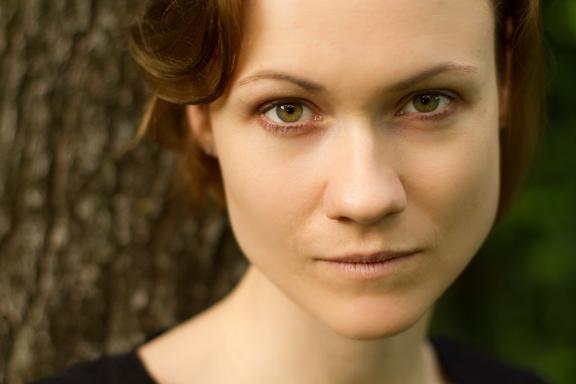}&
   \includegraphics[width=0.32\linewidth]{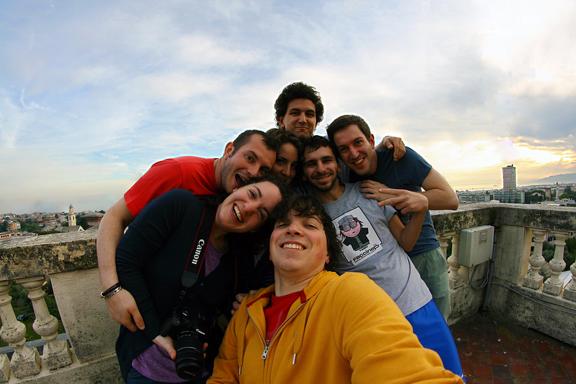}&
   \includegraphics[width=0.32\linewidth]{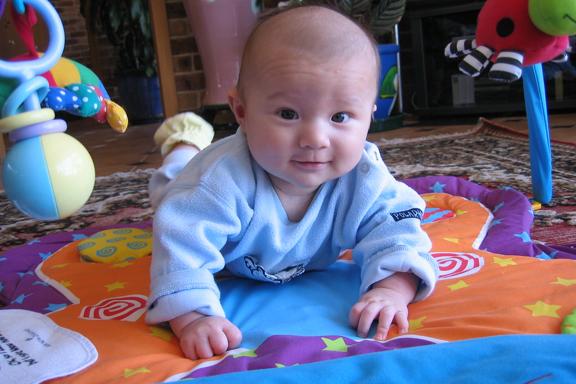}&
   \includegraphics[width=0.32\linewidth]{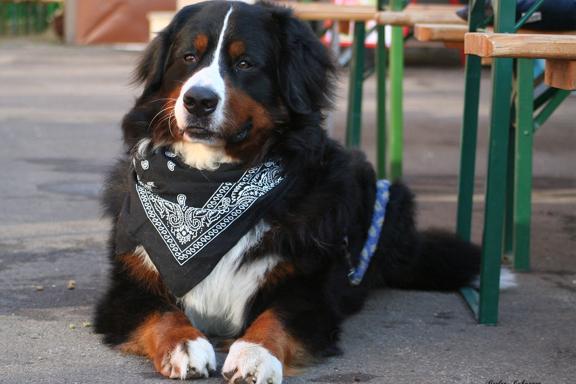}&
   \includegraphics[width=0.32\linewidth]{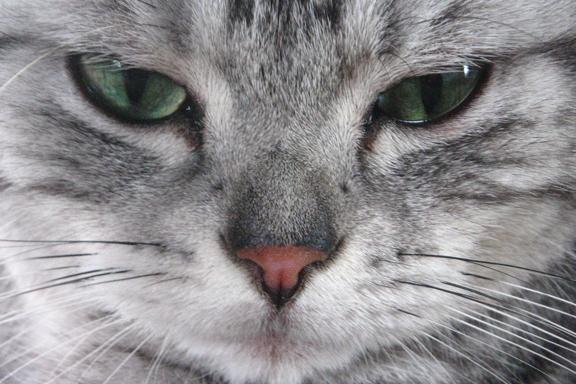}&
   \includegraphics[width=0.32\linewidth]{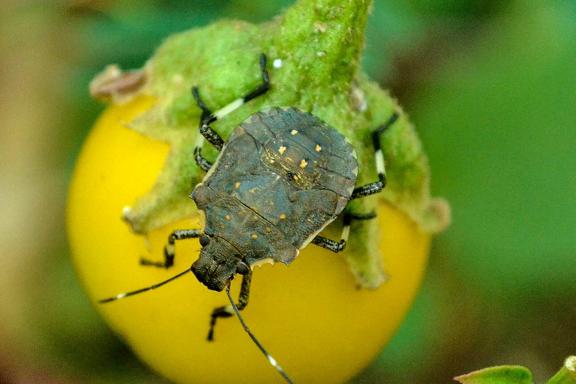}\\
Portrait & Group Portrait & Kids & Dog & Cat & Macro \\
   \includegraphics[width=0.32\linewidth]{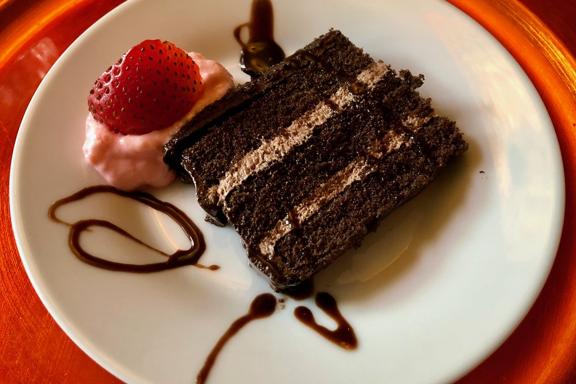}&
   \includegraphics[width=0.32\linewidth]{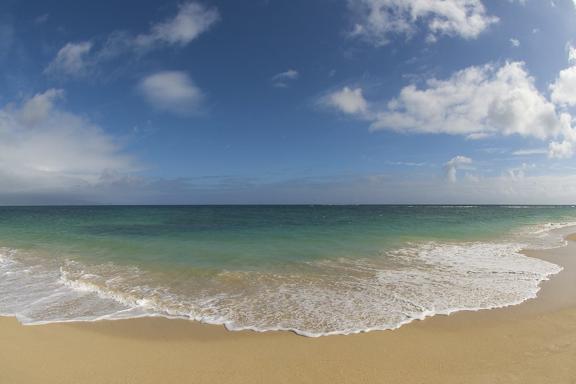}&
   \includegraphics[width=0.32\linewidth]{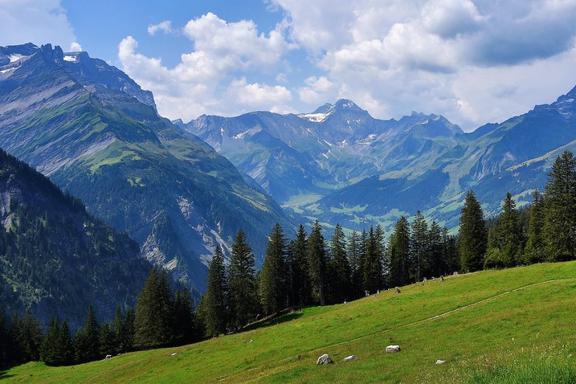}&
   \includegraphics[width=0.32\linewidth]{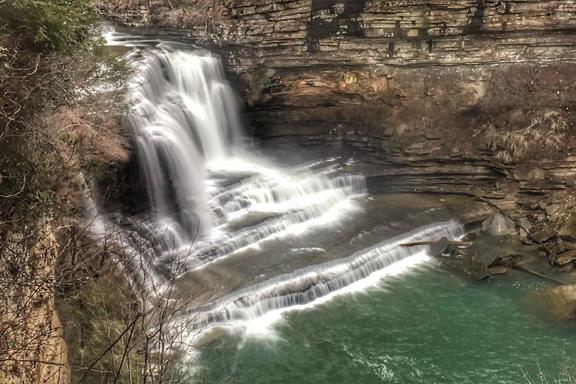}&
   \includegraphics[width=0.32\linewidth]{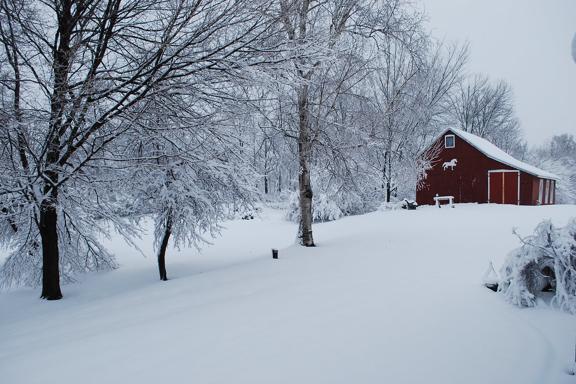}&
   \includegraphics[width=0.32\linewidth]{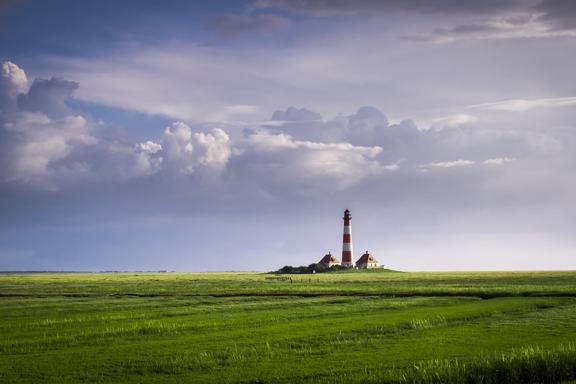}\\
Gourmet & Beach & Mountains & Waterfall & Snow & Landscape \\
   \includegraphics[width=0.32\linewidth]{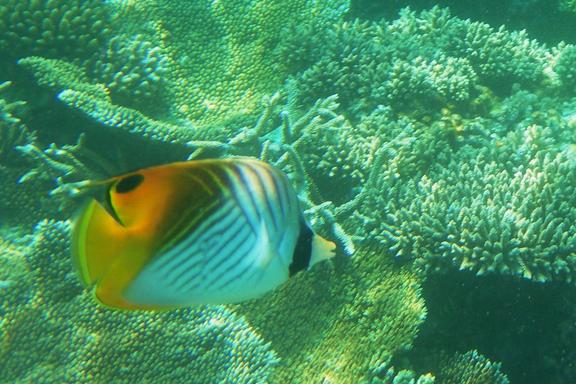}&
   \includegraphics[width=0.32\linewidth]{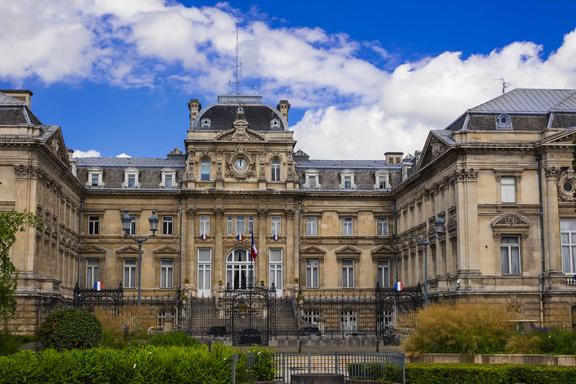}&
   \includegraphics[width=0.32\linewidth]{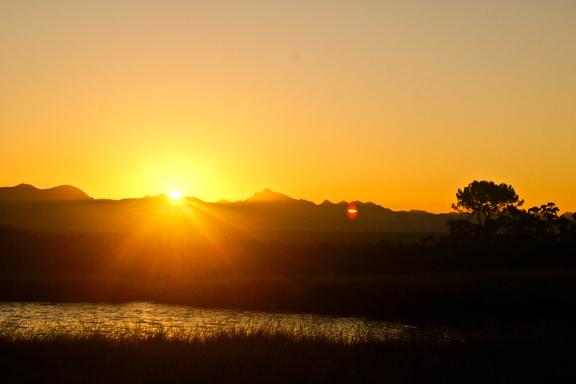}&
   \includegraphics[width=0.32\linewidth]{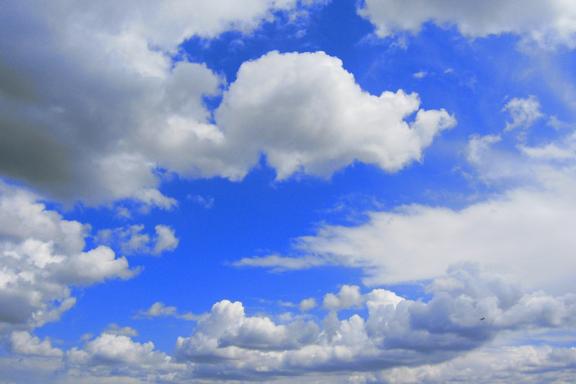}&
   \includegraphics[width=0.32\linewidth]{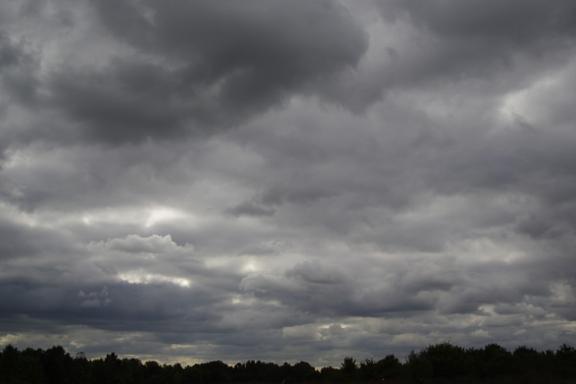}&
   \includegraphics[width=0.32\linewidth]{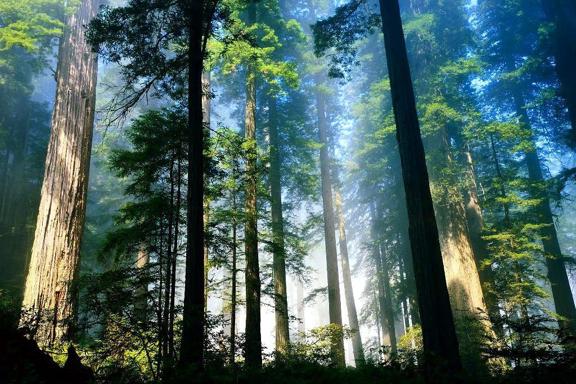}\\
Underwater & Architecture & Sunrise \& Sunset & Blue Sky & Overcast & Greenery \\
   \includegraphics[width=0.32\linewidth]{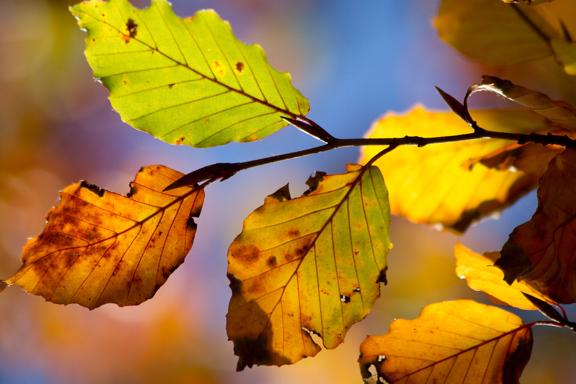}&
   \includegraphics[width=0.32\linewidth]{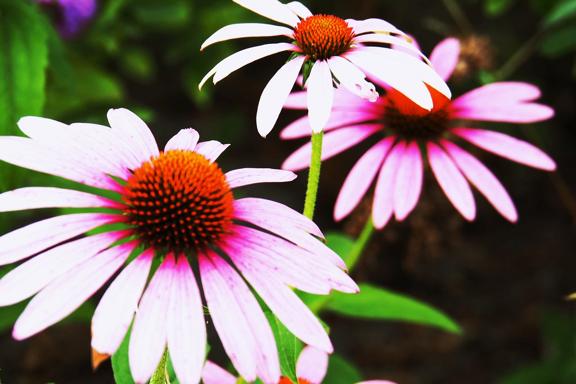}&
   \includegraphics[width=0.32\linewidth]{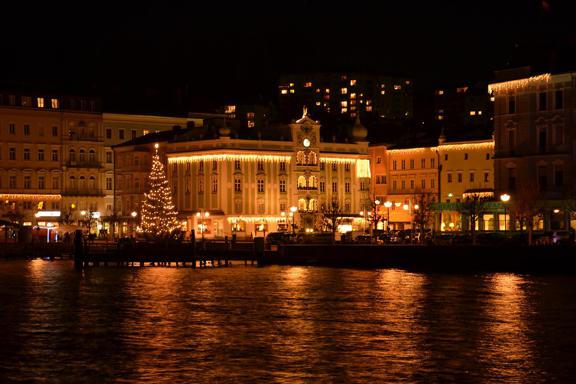}&
   \includegraphics[width=0.32\linewidth]{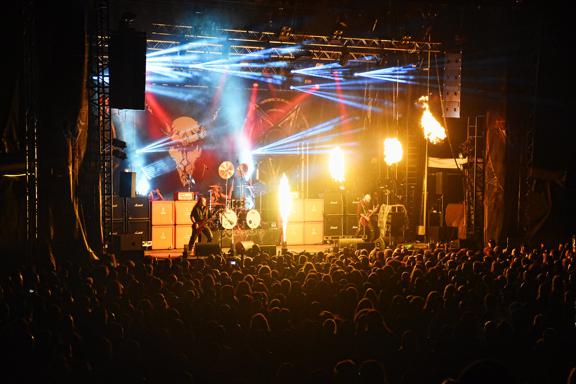}&
   \includegraphics[width=0.32\linewidth]{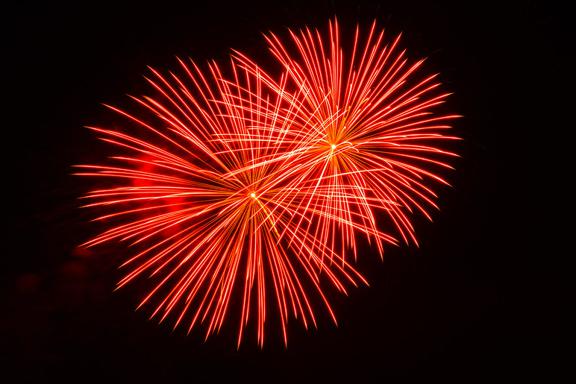}&
   \includegraphics[width=0.32\linewidth]{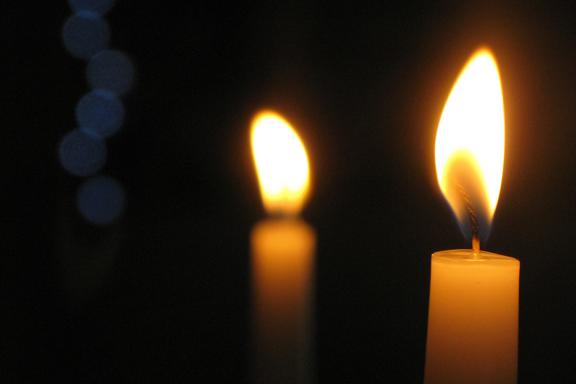}\\
Autumn Plants & Flowers & Night Shot & Stage & Fireworks & Candlelight \\
   \includegraphics[width=0.32\linewidth]{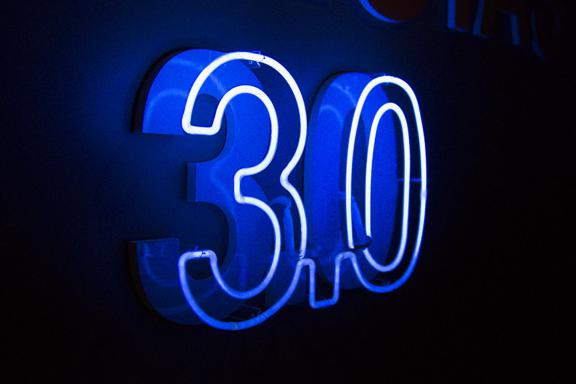}&
   \includegraphics[width=0.32\linewidth]{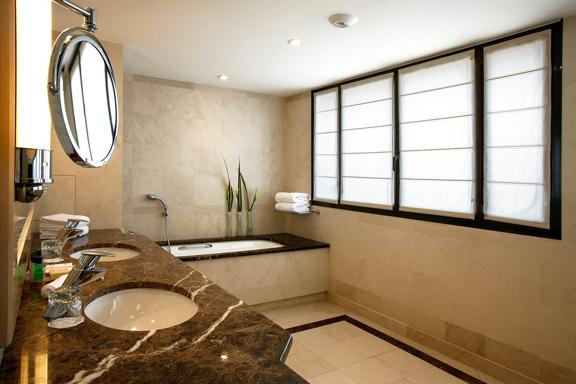}&
   \includegraphics[width=0.32\linewidth]{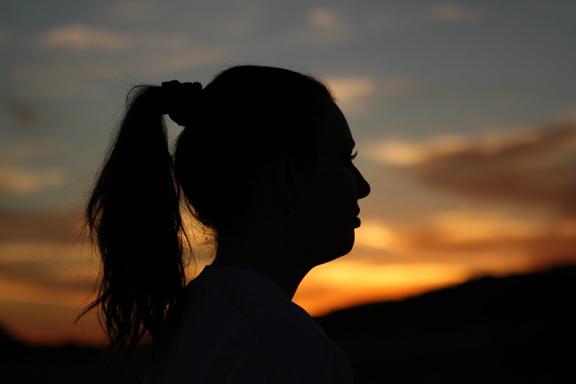}&
   \includegraphics[width=0.32\linewidth]{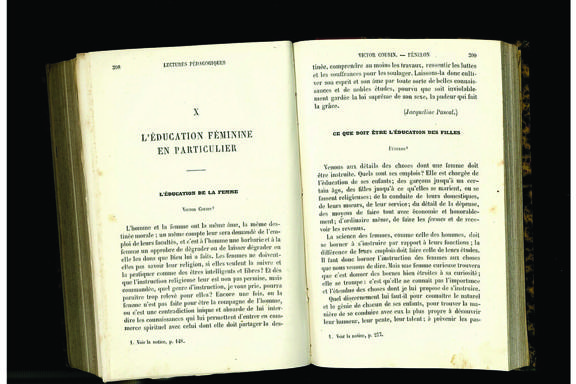}&
   \includegraphics[width=0.32\linewidth]{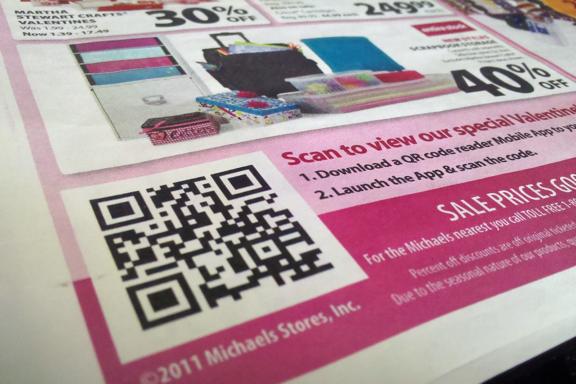}&
   \includegraphics[width=0.32\linewidth]{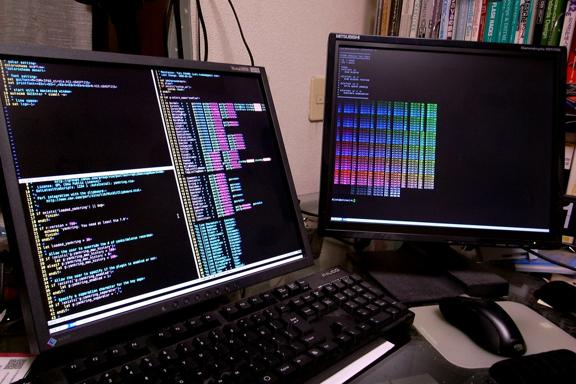}\\
Neon Lights & Indoor & Backlight & Document & QR Code & Monitor Screen \\
\end{tabular}
}
\vspace{2.0mm}
\caption{Visualization of the 30 Camera Scene Detection Dataset (CamSDD) categories.}
\vspace{0.0mm}
\label{fig:dataset}
\end{figure*}

The problem of automatic camera scene prediction on smartphones appeared soon after the introduction of the first mobile cameras. While the initial scene classification approaches were using only manually designed features and some simple machine learning algorithms, the availability of much more powerful AI hardware such as NPUs, GPUs and DSPs made it possible to use considerably more accurate and efficient deep learning-based solutions. Nevertheless, this task has not been properly addressed in the literature until the introduction of the Camera Scene Detection Dataset (CamSDD) dataset in~\cite{pouget2021fast}, where this problem was carefully defined and training data for 30 different camera scene categories was provided along with a fast baseline solution. In this challenge, we take one step further in solving this task by imposing additional efficiency-related constraints on the developed models.

When it comes to the deployment of AI-based solutions on portable devices, one needs to take care of the particularities of mobile CPUs, NPUs and GPUs to design an efficient model. An extensive overview of mobile AI acceleration hardware and its performance is provided in~\cite{ignatov2019ai,ignatov2018ai}. According to the results reported in these papers, the latest mobile NPUs are already approaching the results of mid-range desktop GPUs released not long ago. However, there are still two major issues that prevent a straightforward deployment of neural networks on mobile devices: a restricted amount of RAM, and a limited and not always efficient support for many common deep learning layers and operators. These two problems make it impossible to process high resolution data with standard NN models, thus requiring a careful adaptation of each architecture to the restrictions of mobile AI hardware. Such optimizations can include network pruning and compression~\cite{chiang2020deploying,ignatov2020rendering,li2019learning,liu2019metapruning,obukhov2020t}, 16-bit / 8-bit~\cite{chiang2020deploying,jain2019trained,jacob2018quantization,yang2019quantization} and low-bit~\cite{cai2020zeroq,uhlich2019mixed,ignatov2020controlling,liu2018bi} quantization, device- or NPU-specific adaptations, platform-aware neural architecture search~\cite{howard2019searching,tan2019mnasnet,wu2019fbnet,wan2020fbnetv2}, \etc.

While many challenges and works targeted at efficient deep learning models have been proposed recently, the evaluation of the obtained solutions is generally performed on desktop CPUs and GPUs, making the developed solutions not practical due to the above mentioned issues. To address this problem, we introduce the first \textit{Mobile AI Workshop and Challenges}, where all deep learning solutions are developed for and evaluated on real low-power devices.
In this competition, the participating teams were provided with a large-scale CamSDD dataset containing more than 11K images belonging 30 camera scene categories. Since many mobile and IoT platforms can efficiently accelerate INT8 models only, all developed solutions had to be fully-quantized. Within the challenge, the participants were evaluating the runtime and tuning their models on the Apple Bionic A11 SoC used as the target platform for this task.
The final score of each submitted solution was based on the runtime and top-1 / top-3 accuracy results, thus balancing between the precision and efficiency of the proposed model. Finally, all developed solutions are fully compatible with the TensorFlow Lite framework~\cite{TensorFlowLite2021}, thus can be deployed and accelerated on any mobile platform providing AI acceleration through the Android Neural Networks API (NNAPI)~\cite{NNAPI2021} or custom TFLite delegates~\cite{TFLiteDelegates2021}.

\smallskip


This challenge is a part of the \textit{MAI 2021 Workshop and Challenges} consisting of the following competitions:


\small

\begin{itemize}
\item Learned Smartphone ISP on Mobile NPUs~\cite{ignatov2021learned}
\item Real Image Denoising on Mobile GPUs~\cite{ignatov2021fastDenoising}
\item Quantized Image Super-Resolution on Edge SoC NPUs~\cite{ignatov2021real}
\item Real-Time Video Super-Resolution on Mobile GPUs~\cite{romero2021real}
\item Single-Image Depth Estimation on Mobile Devices~\cite{ignatov2021fastDepth}
\item Quantized Camera Scene Detection on Smartphones
\item High Dynamic Range Image Processing on Mobile NPUs
\end{itemize}

\normalsize


\noindent The results obtained in the other competitions and the description of the proposed solutions can be found in the corresponding challenge papers.


\begin{figure*}[t!]
\centering
\setlength{\tabcolsep}{1pt}
\resizebox{0.96\linewidth}{!}
{
\includegraphics[width=1.0\linewidth]{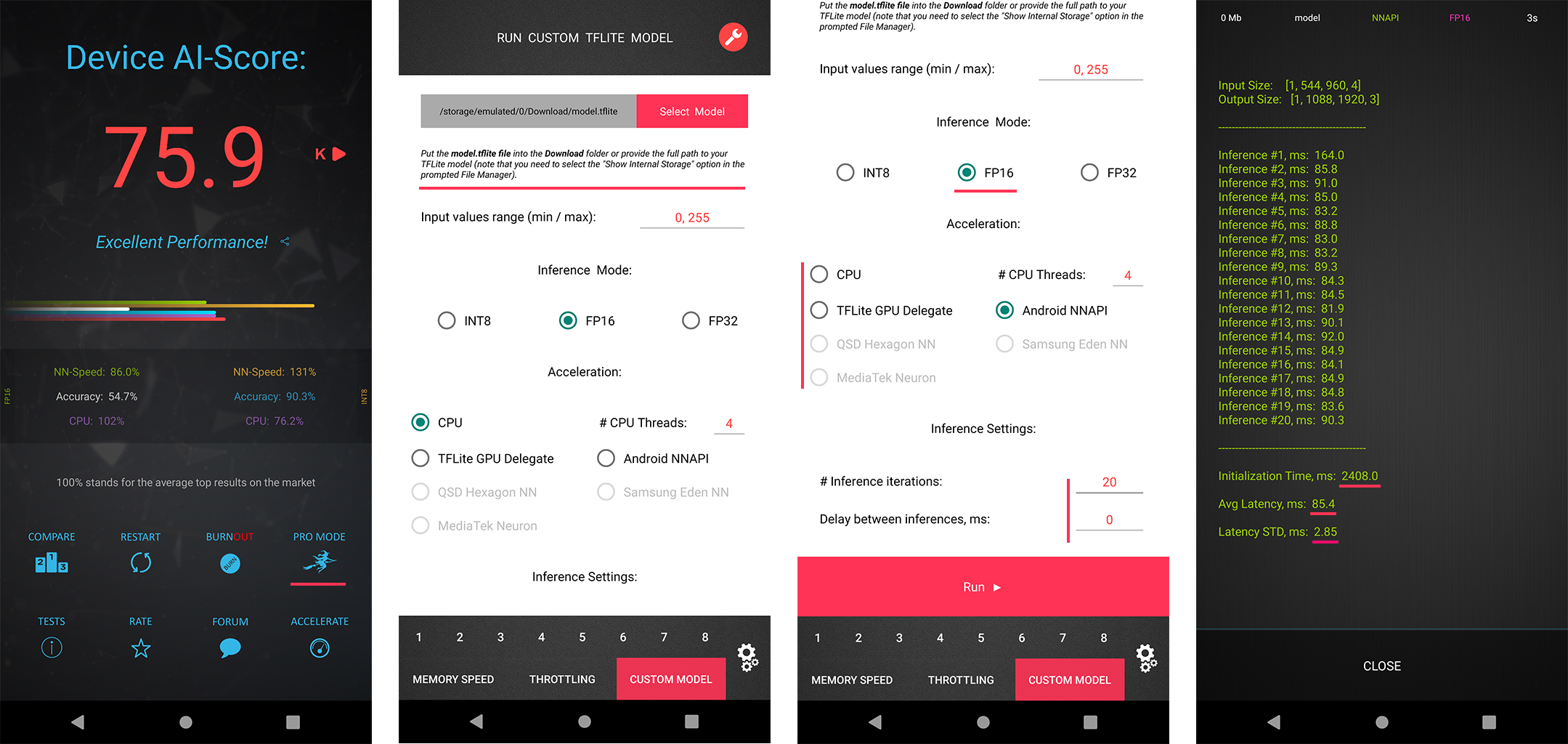}
}
\vspace{0.2cm}
\caption{Loading and running custom TensorFlow Lite models with AI Benchmark application. The currently supported acceleration options include Android NNAPI, TFLite GPU, Hexagon NN, Samsung Eden and MediaTek Neuron delegates as well as CPU inference through TFLite or XNNPACK backends. The latest app version can be downloaded at \url{https://ai-benchmark.com/download}}
\vspace{-1.2mm}
\label{fig:ai_benchmark_custom}
\end{figure*}

\section{Challenge}

To develop an efficient and practical solution for mobile-related tasks, one needs the following major components:

\begin{enumerate}
\item A high-quality and large-scale dataset that can be used to train and evaluate the solution;
\item An efficient way to check the runtime and debug the model locally without any constraints;
\item An ability to regularly test the runtime of the designed neural network on the target mobile platform or device.
\end{enumerate}

This challenge addresses all the above issues. Real training data, tools, and runtime evaluation options provided to the challenge participants are described in the next sections.

\subsection{Dataset}

In this challenge, we use the Camera Scene Detection Dataset (CamSDD)~\cite{pouget2021fast} that provides high-quality diverse data for the considered task. It consists of 30 different scene categories shown in Fig.~\ref{fig:dataset} and contains more than 11K images that were crawled from \textit{Flickr}\footnote{\url{https://www.flickr.com/}} using the same setup as in~\cite{ignatov2018wespe}. All photos were inspected manually to remove monochrome and heavily edited pictures, images with distorted colors and watermarks, photos that are impossible for smartphone cameras (\eg, professional underwater or night shots), etc. The dataset was designed to contain diverse images, therefore each scene category contains photos taken in different places, from different viewpoints and angles: \eg, the ``cat'' category does not only contain cat faces but also normal full-body pictures shot from different positions. This diversity is essential for training a model that is generalizable to different environments and shooting conditions. Each image from the CamSDD dataset belongs to only one scene category. The dataset was designed to be balanced, thus each category contains on average around 350 photos. After the images were collected, they were resized to 576$\times$384 px resolution as using larger photos will not bring any information that is vital for the considered classification problem while will increase the processing time.

\subsection{Local Runtime Evaluation}

\begin{table*}[t!]
\centering
\resizebox{\linewidth}{!}
{
\begin{tabular}{l|c|cc|cc|c|c}
\hline
Team \, & \, Author \, & \, Framework \, & \, Model Size, MB \, & \, Top-1, \%	& Top-3, \%  & Final Runtime, ms & Final Score \\
\hline
\hline
ByteScene & halfmoonchen              & Keras / TensorFlow &   8.2  & 95.00  & 99.50 &   4.44    & 163.08 \\
EVAI & EVAI                           & Keras / TensorFlow &   0.83 & 93.00  & 98.00 &   3.35    & 19.1 \\
\rowcolor{grayhighlight} \textit{MobileNet-V2}~\cite{pouget2021fast} & \textit{Baseline}
                                      & TensorFlow         &   18.6 & 94.17  & 98.67 &  16.38    & 13.99 \\
ALONG & YxChen03                      & TensorFlow         &   12.7 & 94.67  & 99.50 &  64.45    & 8.94 \\
Team Horizon & tensorcat              & Keras / TensorFlow &   2.27 & 92.33  & 98.67 &   7.7     & 8.31 \\
Airia-Det & stvea                     & TensorFlow         &   1.36 & 93.00  & 99.00 &  17.51    & 7.31 \\
DataArt Perceptrons & andrewBatutin   & Keras / TensorFlow &   2.73 & 91.50  & 97.67 &  54.13    & 0.33 \\
PyImageSearch & thanki                & TensorFlow         &   2.02 & 89.67  & 97.83 &  45.88    & 0.12 \\
neptuneai & neptuneai                 & Keras / TensorFlow &  0.045 & 83.67  & 94.67 &   4.17    & 0 \\
Sidiki & Sidiki                       & Keras / TensorFlow &  0.072 & 78.00  & 93.83 &   1.74    & 0 \\
\end{tabular}
}
\vspace{2.6mm}
\caption{\small{Mobile AI 2021 Quantized Camera Scene Detection challenge results and final rankings. The runtime values were obtained on 576$\times$384 px images on the Apple Bionic A11 platform. Top-1 / Top-3 accuracy was computed using the provided fully-quantized TFLite models. Team \textit{ByteScene} is the challenge winner, \textit{Baseline} corresponds to the MobileNet-V2 based solution presented in~\cite{pouget2021fast}.}}
\label{tab:results}
\vspace{-1.2mm}
\end{table*}

When developing AI solutions for mobile devices, it is vital to be able to test the designed models and debug all emerging issues locally on available devices. For this, the participants were provided with the \textit{AI Benchmark} application~\cite{ignatov2018ai,ignatov2019ai} that allows to load any custom TensorFlow Lite model and run it on any Android device with all supported acceleration options. This tool contains the latest versions of \textit{Android NNAPI, TFLite GPU, Hexagon NN, Samsung Eden} and \textit{MediaTek Neuron} delegates, therefore supporting all current mobile platforms and providing the users with the ability to execute neural networks on smartphone NPUs, APUs, DSPs, GPUs and CPUs.

\smallskip

To load and run a custom TensorFlow Lite model, one needs to follow the next steps:

\begin{enumerate}
\setlength\itemsep{0mm}
\item Download AI Benchmark from the official website\footnote{\url{https://ai-benchmark.com/download}} or from the Google Play\footnote{\url{https://play.google.com/store/apps/details?id=org.benchmark.demo}} and run its standard tests.
\item After the end of the tests, enter the \textit{PRO Mode} and select the \textit{Custom Model} tab there.
\item Rename the exported TFLite model to \textit{model.tflite} and put it into the \textit{Download} folder of the device.
\item Select mode type \textit{(INT8, FP16, or FP32)}, the desired acceleration/inference options and run the model.
\end{enumerate}

\noindent These steps are also illustrated in Fig.~\ref{fig:ai_benchmark_custom}.

\subsection{Runtime Evaluation on the Target Platform}

In this challenge, we use the \textit{Apple Bionic A11} chipset with the 1-st generation \textit{Neural Engine} that can be found in the iPhone X and iPhone 8 / 8 Plus smartphones as our target runtime evaluation platform. The runtime of all solutions was tested using the TensorFlow Lite CoreML delegate~\cite{TFLite2021CoreML} containing many important performance optimizations for this SoC, the latest iOS 14 operating system was installed on the smartphone. Within the challenge, the participants were able to upload their TFLite models to the runtime validation server connected to a real iPhone device and get instantaneous feedback: the runtime of their solution or an error log if the model contains some incompatible operations. The same setup was also used for the final runtime evaluation.

\subsection{Challenge Phases}

The challenge consisted of the following phases:

\vspace{-0.8mm}
\begin{enumerate}
\item[I.] \textit{Development:} the participants get access to the data and AI Benchmark app, and are able to train the models and evaluate their runtime locally;
\item[II.] \textit{Validation:} the participants can upload their models and predictions to the remote server to check the accuracy on the validation dataset, to get the runtime on the target platform, and to compare their results on the validation leaderboard;
\item[III.] \textit{Testing:} the participants submit their final results, codes, TensorFlow Lite models, and factsheets.
\end{enumerate}
\vspace{-0.8mm}

\subsection{Scoring System}

\begin{table*}[t!]
\centering
\resizebox{\linewidth}{!}
{
\begin{tabular}{l|cc|cc|cc|cc}
\hline
Mobile SoC & \, Snapdragon 888 \, & \,  Snapdragon 855 \,  & \,  Dimensity 1000+ \,  & \,  Dimensity 800 \,  & \,  Exynos 2100 \,  & \, Exynos 990 \, & \, Kirin 990 5G \, & \, Kirin 980 \, \\
AI Accelerator & \, \small Adreno 660 GPU, fps \, & \, \small Hexagon 690, fps \, & \, \small APU 3.0 (6 cores), fps \, & \, \small APU 3.0 (4 cores), fps \, & \, \small NPU, fps \, & \, \small Mali-G77 GPU, fps \, & \, \small Mali-G76 GPU, fps \, & \, \small Mali-G76 GPU, fps \, \\
\hline
\hline
ByteScene           & 198 & 191 & 224 & 156 & \textbf{262} & 98  & 88  & 94  \\
EVAI                & 290 & 347 & \textbf{408} & 275 & 292 & 144 & 138 & 166 \\
ALONG               & 53  & 53  & \textbf{71}  & 52  & 37  & 27  & 27  & 30  \\
Team Horizon        & \textbf{244} & 224 & 171 & 122 & 174 & 101 & 111 & 110 \\
Airia-Det           & \textbf{207} & 126 & 190 & 125 & 92  & 68  & 64  & 74  \\
DataArt Perceptrons & 91  & 68  & 61  & 41  & 30  & 28  & \textbf{92}  & 42  \\
PyImageSearch       & \textbf{94}  & 68  & 73  & 51  & 39  & 32  & 51  & 43  \\
neptuneai           & \textbf{578} & 274 & 252 & 144 & 204 & 181 & 195 & 245 \\
Sidiki              & \textbf{427} & 400 & 198 & 161 & 373 & 228 & 164 & 154 \\
\end{tabular}
}
\vspace{2.6mm}
\caption{\small{The speed of the proposed solutions on several popular mobile platforms. The runtime was measured with the AI Benchmark app using the fastest acceleration option for each device. The best FPS rate for each solution is denoted in bold.}}
\label{tab:runtime_results}
\end{table*}

All solutions were evaluated using the following metrics:

\vspace{-0.8mm}
\begin{itemize}
\setlength\itemsep{-0.2mm}
\item Top-1 accuracy checking how accurately was recognized the main scene category,
\item Top-3 accuracy assessing the quality of the top 3 predicted classes,
\item The runtime on the target Apple Bionic A11 platform.
\end{itemize}
\vspace{-0.8mm}

The score of each final submission was evaluated based on the next formula ($C$ is a constant normalization factor):

\smallskip
\begin{equation*}
\text{Final Score} \,=\, \frac{2^{\text{\, (Top-1} + \text{Top-3) \,}}}{C \cdot \text{runtime}},
\end{equation*}
\smallskip

During the final challenge phase, the participants did not have access to the test dataset. Instead, they had to submit their final TensorFlow Lite models that were subsequently used by the challenge organizers to check both the runtime and the accuracy results of each submission under identical conditions. This approach solved all the issues related to model overfitting, reproducibility of the results, and consistency of the obtained runtime/accuracy values.

\section{Challenge Results}

From above 120 registered participants, 10 teams entered the final phase and submitted valid results, TFLite models, codes, executables and factsheets. Table~\ref{tab:results} summarizes the final challenge results and reports Top-1 / Top-3 accuracy and the runtime numbers for each submitted solution on the final test dataset and on the target evaluation platform. The proposed methods are described in section~\ref{sec:solutions}, and the team members and affiliations are listed in Appendix~\ref{sec:apd:team}.

\subsection{Results and Discussion}

All submitted solutions demonstrated a very high efficiency: the majority of models are able to detect the camera scene category at more than 50 FPS on the target Bionic A11 SoC while achieving a top-3 accuracy of over 97-98\%. All teams were using either MobileNet~\cite{sandler2018mobilenetv2,howard2019searching} or EfficientNet~\cite{tan2019efficientnet} backbones pre-trained on the ImageNet dataset except for team \textit{Sidiki} that proposed a very tiny classification model which size is only 73KB. Despite using only 60K parameters, it is still able to achieve a top-3 accuracy of more than 93\% on the final test set, thus might be practically useful for constraint embedded systems with very tough requirements regarding the model size and memory consumption. Team \textit{ByteScene} is the challenge winner~--- the proposed MobileNet-V3 based model was able to outperform all other solutions in terms of top-1 accuracy while showing more than 220 FPS on the Bionic SoC. To achieve these results, the authors used the Hierarchical Neural Architecture Search to adapt the baseline architecture to the target accuracy / runtime constraints, and additional training data obtained by pseudo-labeling with a larger and more accurate network pre-trained on the CamSDD dataset. The same approach of data augmentation was also used by team \textit{ALONG} that generated additional 50K training images using a similar strategy.

To further benchmark the efficiency of the designed solutions, we additionally tested their performance on several popular smartphone chipsets. The runtime results demonstrated in Table~\ref{tab:runtime_results} were measured with the AI Benchmark using the most efficient acceleration option. We should note that though the Snapdragon 888 and the Kirin 990 5G SoCs have NPUs / DSPs supporting quantized inference, they are not compatible with TensorFlow-2.x models right now, thus in these two cases their GPUs were used instead for accelerating the networks. As one can see, all models were able to achieve real-time performance with more than 27-30 classified images per second on all considered platforms. The winning solution from team \textit{ByteScene} was able reach more than 150-200 FPS on the latest chipsets from Qualcomm, MediaTek and Samsung, and should be able to perform real-time classification on all recent high-end, mid-range and even low-end platforms. The solution from team \textit{EVAI} can provide an additional 50-100\% runtime boost at the expense of a slightly lower accuracy. Overall, the obtained results also demonstrate the efficiency of mobile AI accelerators for image classification tasks as they can achieve enormous processing rates while maintaining low power consumption.

\section{Challenge Methods}
\label{sec:solutions}

\noindent This section describes solutions submitted by all teams participating in the final stage of the MAI 2021 Real-Time Camera Scene Detection challenge.

\subsection{ByteScene}

\begin{figure}[h!]
\centering
\resizebox{1.0\linewidth}{!}
{
\includegraphics[width=1.0\linewidth]{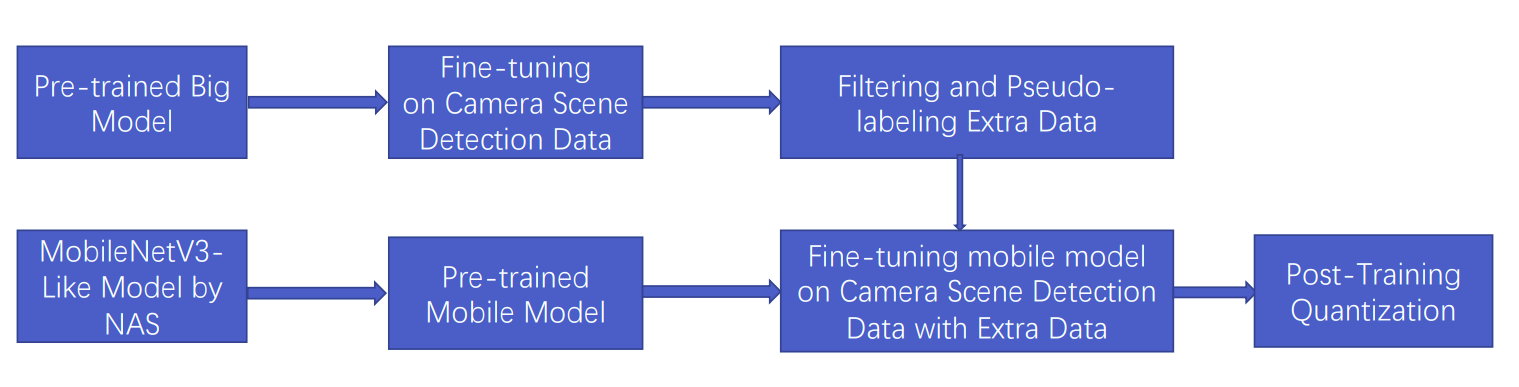}
}
\vspace{0mm}
\caption{\small{The training scheme used by ByteScene team.}}
\label{fig:bytescene_diagram}
\vspace{-2mm}
\end{figure}

Team ByteScene used a transfer learning method (Fig.~\ref{fig:bytescene_diagram}) inspired by the Big Transfer (BiT)~\cite{kolesnikov2019big} to train mobile and big classification models. The first one is using a MobileNet-V3 like architecture demonstrated in Table~\ref{tab:bytescene_architecture} that was searched using the Hierarchical Neural Architecture Search~\cite{xia2020hnas} with additional runtime constraints, and achieves a top-1 accuracy of 67.82\% on the ImageNet2012 validation set. Two fully-connected layers were appended on top of the model to the 86M FLOPs backbone. To reduce the amount of computations, the input image was resized from 576$\times$384px to 128$\times$128px, normalized to range $[-1, +1]$, and then passed to the network.

\textBF{Big model training.} The authors based their large model on the ResNet101x3 architecture. With 1003 held-out labeled images as a validation set, it was first fine-tuned on the CamSDD data. During fine-tuning, its backbone was frozen and only classifier parameters were optimized for 10 epochs using AdamW~\cite{loshchilov2017decoupled} with a batch size of 256, an initial learning rate of $1.5e-3$ and a piece-wise constant decay by a factor of 10 at steps of 200 and 300. Sparse categorical cross-entropy was used as an objective function. During the training process, the images were resized to 256$\times$256px and then randomly cropped to 224$\times$224px resolution. After performing pseudo-labeling of extra data to get a new validation set, the model was trained again using all available training images. The resulting big model achieved a top-1 accuracy of 97.83\% on the official validation set.

\textBF{Mobile model training.} The pre-trained on the ImageNet2012 mobile model (Table~\ref{tab:bytescene_architecture}) was first fine-tuned on the CamSDD data with a frozen backbone using AdamW optimizer with an initial learning rate of $1.5e-3$, a weight decay of $4e-5$, a batch size 256, and a schedule length of 400. The objective function was the same as for training of the big model. The input images were resized to 160$\times$160px and then randomly cropped to 128$\times$128px. Next, the backbone was unfreezed, and the model was fine-tuned for additional 800 steps with the same optimization parameters. Finally, the backbone was frozen again, and the model was fine-tuned for additional 400 steps with the SGDW optimizer. During the third fine-tuning, the training images were directly resized to 128$\times$128 pixels. In addition to the official 9897 CamSDD training images, 2577 extra training images were used (pseudo-labeled by the pre-trained big model) when training the mobile model.

The resulting quantized INT8 TFLite model was obtained using the standard TensorFlow's post-training quantization tools. To preserve the model accuracy after quantization, the authors used only ReLU6 and HardSigmoid~\cite{howard2019searching} nonlinearities.

\begin{table}[t!]
\centering
\resizebox{0.97\linewidth}{!}
{
\begin{tabular}{lcccc}
\hline
Input shape  & Block & Exp Size &SE & Stride  \\
\hline
\(128^2 \times 3\)  & conv2d, $3 \times 3$  &16   &No   & 2      \\
\(64^2 \times 16\)  & bneck, $3 \times 3$   &16   &No   & 1     \\
\(64^2 \times 16\)  & bneck, $3 \times 3$   &48   &Yes  & 2   \\
\(32^2 \times 24\)  & bneck, $3 \times 3$   &72   &Yes   & 2   \\
\(16^2 \times 32\)  & bneck, $3 \times 3$   &64   &Yes   & 1   \\
\(16^2 \times 32\)  & bneck, $3 \times 3$   &96   &Yes   & 1   \\
\(16^2 \times 32\)  & bneck, $3 \times 3$   &96   &No   & 2   \\
\(8^2 \times 64\)   & bneck, $3 \times 3$   &128   &No   & 1   \\
\(8^2 \times 64\)   & bneck, $3 \times 3$   &256   &No   & 1   \\
\(8^2 \times 64\)   & bneck, $3 \times 3$   &320   &Yes   & 1   \\
\(8^2 \times 96\)   & bneck, $3 \times 3$   &192   &Yes   & 1   \\
\(8^2 \times 96\)   & bneck, $3 \times 3$   &288   &Yes   & 1   \\
\(8^2 \times 96\)   & bneck, $3 \times 3$   &576   &Yes   & 2   \\
\(4^2 \times 192\)   & bneck, $3 \times 3$   &768   &Yes   & 1   \\
\(4^2 \times 192\)   & bneck, $3 \times 3$   &960   &Yes   & 1   \\
\(4^2 \times 192\)   & bneck, $3 \times 3$   &768   &Yes   & 1   \\
\(4^2 \times 192\)   & bneck, $3 \times 3$   &960   &Yes   & 1   \\
\(4^2 \times 192\)   & bneck, $3 \times 3$   &960   &Yes   & 1   \\
\(4^2 \times 192\)   & bneck, $3 \times 3$   &768   &Yes   & 1   \\
\(4^2 \times 192\)   & bneck, $3 \times 3$   &1152   &Yes   & 1   \\
\(4^2 \times 192\)  & conv2d, $1 \times 1$  &1024      &No   & 1       \\
\(4^2 \times 1024\) &global avgpool    &1024     &No   & 1       \\
\(1^2 \times1024\)  &fc, Relu6    & 1280  &No   & -       \\
\(1280\)    &fc    & 30  &No   & -       \\
\hline
\end{tabular}
}
\vspace{2.2mm}
\caption{The model architecture proposed by ByteScene. The MobilenetV3 Bottleneck blocks \textit{(bneck)}~\cite{howard2019searching} are used to build the model. For a \textit{bneck} block, \textit{Exp Size} denotes the number of channels in the expansion layer. For the other blocks, \textit{Exp Size} denotes the output channel number of a block. \textit{SE} denotes whether there is a \textit{Squeeze-And-Excite} op in the block.}
\label{tab:bytescene_architecture}
\vspace{-2mm}
\end{table}

\subsection{EVAI (Edge Vision and Visual AI)}

\begin{figure}[h!]
\centering
\resizebox{1.0\linewidth}{!}
{
\includegraphics[width=1.0\linewidth]{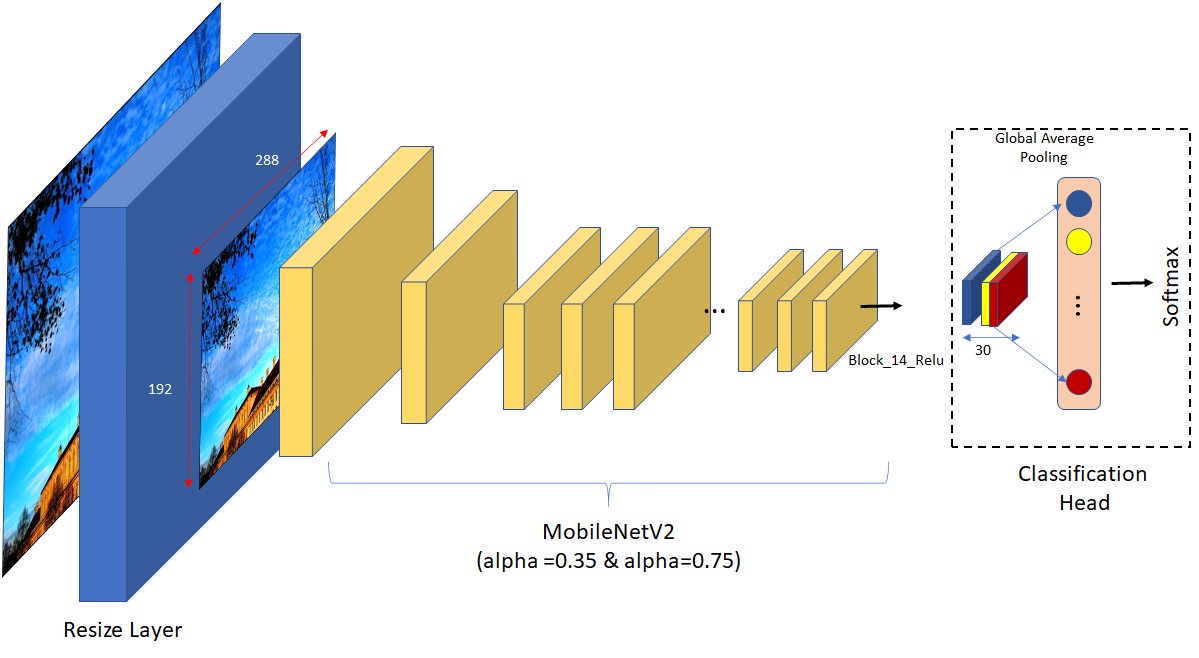}
}
\vspace{1.2mm}
\caption{\small{The model architecture proposed by team EVAI.}}
\label{fig:evai_plot}
\vspace{-2mm}
\end{figure}

Team EVAI developed a MobileNetV2-like~\cite{sandler2018mobilenetv2} model with $alpha=0.75$ and an input shape of 96$\times$144px. This solution demonstrated the best accuracy-latency tradeoff according the experiments conducted by the authors (Table~\ref{tab:evai_models_comp}).

The architecture of the final model is shown in Fig.~\ref{fig:evai_plot}. A resize layer is introduced to reduce the resolution if the input images by 4 times. The last three blocks (\textit{block15}, \textit{block16} and the last convolutional block) of the original MobileNet-V2 model were removed to reduce the computations. The classification head contains a convolutional block with separable convolution layer, batch normalization layer and \textit{ReLU} activations. The output of this block has 30 feature maps, its spacial resolution is then reduced with one global average pooling layer, and the final predictions are obtained after applying the \textit{Softmax} activation. The proposed network does not have any dense layers.

The model was trained for 200 epochs to minimize categorical cross-entropy loss function with label smoothing $\epsilon=0.1$. The model parameters were optimized using stochastic gradient descent with momentum $0.9$ and exponential learning rate scheduler with the initial learning rate of $5e-2$. The authors used a batch size is $128$ and applied additional data augmentation during the training such as flips, grid distortions, shift-scale-rotations, perspective transforms, random resized crops, color jitter and blurring. The final INT8 model was obtained using TensorFlow's post-training quantization tools.

\begin{table}[ht]
\centering
\resizebox{0.99\columnwidth}{!}{
\begin{tabular}{|c c c c c|}
\hline
Model Details & \, Val. Accuracy \, & \, Val. Accuracy \, & \, Latency \, & \, Latency \, \\ [0.5ex]
& FP32, \% & INT8, \%  & FP32, ms & INT8, ms \\
\hline
\hline		
\, MobileNetV2 a=0.35 \, &   &  &  &  \\ 	
$96\times144$ & 90.3  & 90 & 8 & 3 \\
MobileNetV2 a=0.35 &   &  &  &  \\ 	
$192\times288$ & 95 & 94.5  & 14 & 8 \\
MobileNetV2 a=0.75 &   &  &  &  \\ 	
$96\times144$ & 93.6  & 92.6 & 9 & 4 \\
\hline
\end{tabular}
}
\vspace{2.2mm}
\caption{The performance of different MobileNet-V2 backbones obtained by team EVAI.}
\label{tab:evai_models_comp}
\end{table}

\subsection{ALONG}

\begin{figure}[h!]
\centering
\resizebox{1.0\linewidth}{!}
{
\includegraphics[width=1.0\linewidth]{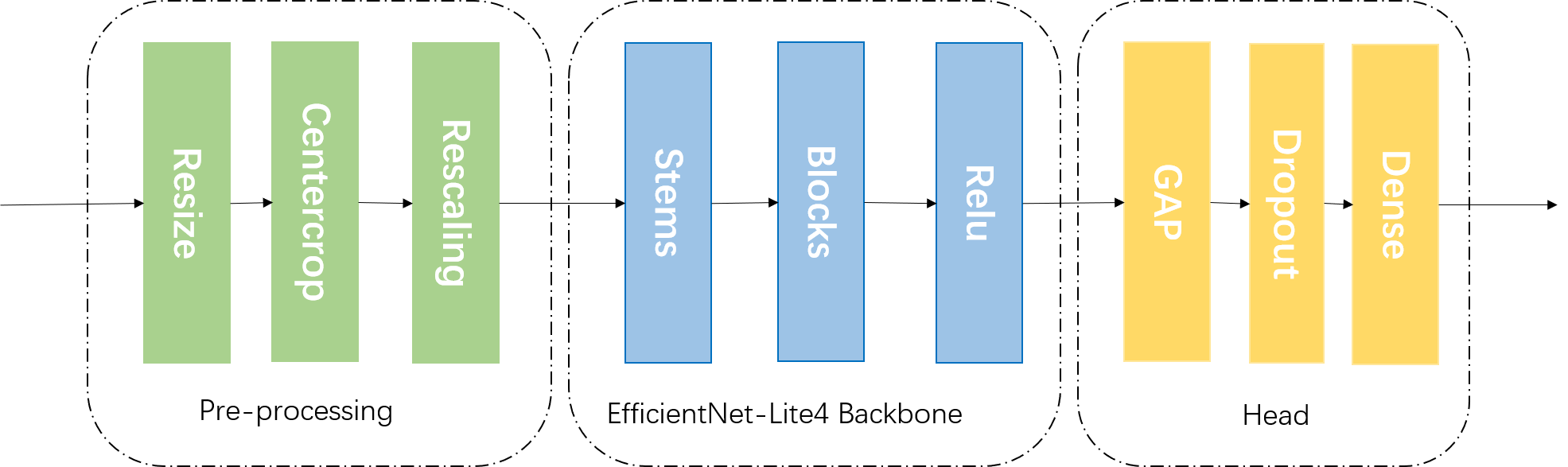}
}
\vspace{1.2mm}
\caption{\small{The EfficientNet-based model proposed by ALONG.}}
\label{fig:along_plot}
\end{figure}

Team ALONG based its solution on the EfficientNet-Lite4~\cite{tan2019efficientnet} model, the architecture of the model is presented in Fig.~\ref{fig:along_plot}. The authors used web crawlers to collect a large number of additional training images. Self-distillation method with 4-fold cross-validation was used to retain images with only high classification probability~--- 50K images were retained and integrated with the original dataset. This extra data improved the classification accuracy by around 2\%. An additional 1\% accuracy increase was achieved by re-weighting misclassified samples. Since using the original images for model training can result in overfitting, the authors used various data augmentation methods including random cropping, rotation, flipping, brightness and contrast adjustments, blur and mix-cut. In addition, dropout and weight decay were applied. The classification accuracy of the model was improved by around 1\% through the above methods (Table~\ref{tab:along_results}).

The model parameters were optimized using SGD with an initial learning rate of 0.01 and momentum 0.9 for 200 epochs with a batch size of 320. The learning rate scheduler was as follows: 0.01 for the first 2 epochs, 0.001 for the next 118 epochs, and 0.0001 for the rest of the training. The objective function is the categorical cross-entropy with label smoothing ($\epsilon=0.2$). The final INT8 model was obtained using TensorFlow's post-training quantization tools.

\begin{table}[ht]
\centering
\label{tab:along_results}
\resizebox{0.99\columnwidth}{!}{
    	\begin{tabular}{|c c c c c c|}
        \hline
        Method & QPT & QAT & Improvement & Acc & Latency\\
        \hline
        NasNetMobile & + & + & - & 92.1\%  &40ms\\
        MobileNetV3Large & + & -  & - & 93.3\%  &77ms\\
        EfficientNet-B1 & + & -  & - & 93.6\%  &89ms\\
        NasNetLarge & - & -  &  - & 96.6\%
        &1500ms\\
        Bit-m R50x1 & - & -  & +ImageNet21K Pretrained & 97.8\%
        &1200ms\\
        \hline
        EfficientNet-Lite0 & + & +  & - & 92.0\% &33ms\\
        EfficientNet-Lite4 & + & +  & - & 93.0\% &74ms\\
        EfficientNet-Lite4 & + & +  & + Data Augmentation & 94.0\% &74ms\\
        EfficientNet-Lite4 & + & +  & + Image Collection & 95.0\% &74ms\\
        EfficientNet-Lite4 & + & +  & + Samples Reweighting & 94.0\% &74ms\\
        EfficientNet-Lite4 & + & +  & + All & 97.0\% &74ms\\
        \hline
        \end{tabular}
}
\vspace{2mm}
\caption{The performance of different architectures and training strategies obtained by team ALONG on the validation set. \textit{QPT} stands for quantization post training, and \textit{QAT} means quantization-aware training.}
\end{table}

\subsection{Team Horizon}

The authors used a MobileNetV2-based model, where the input image was resizes to $1/3$ of its original resolution to improve the runtime results at the expense of minor reduction in accuracy. The model was trained in 4 steps. First, the parameters of the classification head were optimized with Adam for 10 epochs with the default learning rate, then the classification head was trained for another 10 epochs with a reduced learning rate of $1e-4$. Next, the whole model was trained for 10 epochs with a learning rate of $1e-5$, and at the last step it was optimized again with a learning rate of $1e-6$. The network was trained with a batch size of 256 to minimize the cross-entropy loss function in all steps.

\subsection{Airia-Det}

Team Airia-Det also based its solution on the MobileNetV2~\cite{sandler2018mobilenetv2} architecture with several modifications. In the classification head, a 5$\times$5 depth separable convolution is used instead of two 3$\times$3 convolutional layers to obtain a larger receptive field and to reduce the number of parameters. The input of the model is resized from 576$\times$384px to 384$\times$384px. The network was trained to minimize the categorical cross-entropy with label smoothing ($\epsilon=0.1$) using Adam optimizer and a batch size of 32.

\subsection{DataArt Perceptrons}

\begin{table}[ht]
\centering
\label{tab:DataArt_results}
\resizebox{0.99\columnwidth}{!}{
\begin{tabular}[]{@{}lccccl@{}}
TFLite Model & Type & Size & Top-1 & iPhone 12 Pro & Score \\
& & (MB) & (\%) & CPU, ms & \\
\hline
\hline
MobileNet-V2 & FP & 8.6 & 92.5 & 83 & 0.0941 \\
MobileNet-V2 & INT8 & 2.7 & 91.3 & 58 & 0.0255 \\
EfficientNet & FP & 29.5 & 96 & 270 & 3.7037 \\
EfficientNet & INT8 & 9.1 & 95.3 & 280 & 1.3533 \\
\hline
\end{tabular}
}
\vspace{1.2mm}
\caption{The performance on the validation dataset and the runtime of different models developed by DataArt Perceptrons team.}
\end{table}

The authors used a standard MobileNet-V2 model and trained it on the provided images with various data augmentation methods such as rescaling, random rotations, flips, contrast adjustment and translation. The size of the input layer shape was set to (384, 576, 3) to fit the challenge requirements, the layers of the backbone model were freezed during the training. The final model was trained using Adam optimizer with a learning rate of $5e-3$, a batch size 32 and the categorical cross-entropy loss function. Though the authors obtained better results with the EfficientNet model (Table~\ref{tab:DataArt_results}), they faced the problems related to its quantization and decided to use the MobileNetV2-based solution instead.

\subsection{PyImageSearch}

\begin{figure}[h!]
\centering
\resizebox{1.0\linewidth}{!}
{
\includegraphics[width=1.0\linewidth]{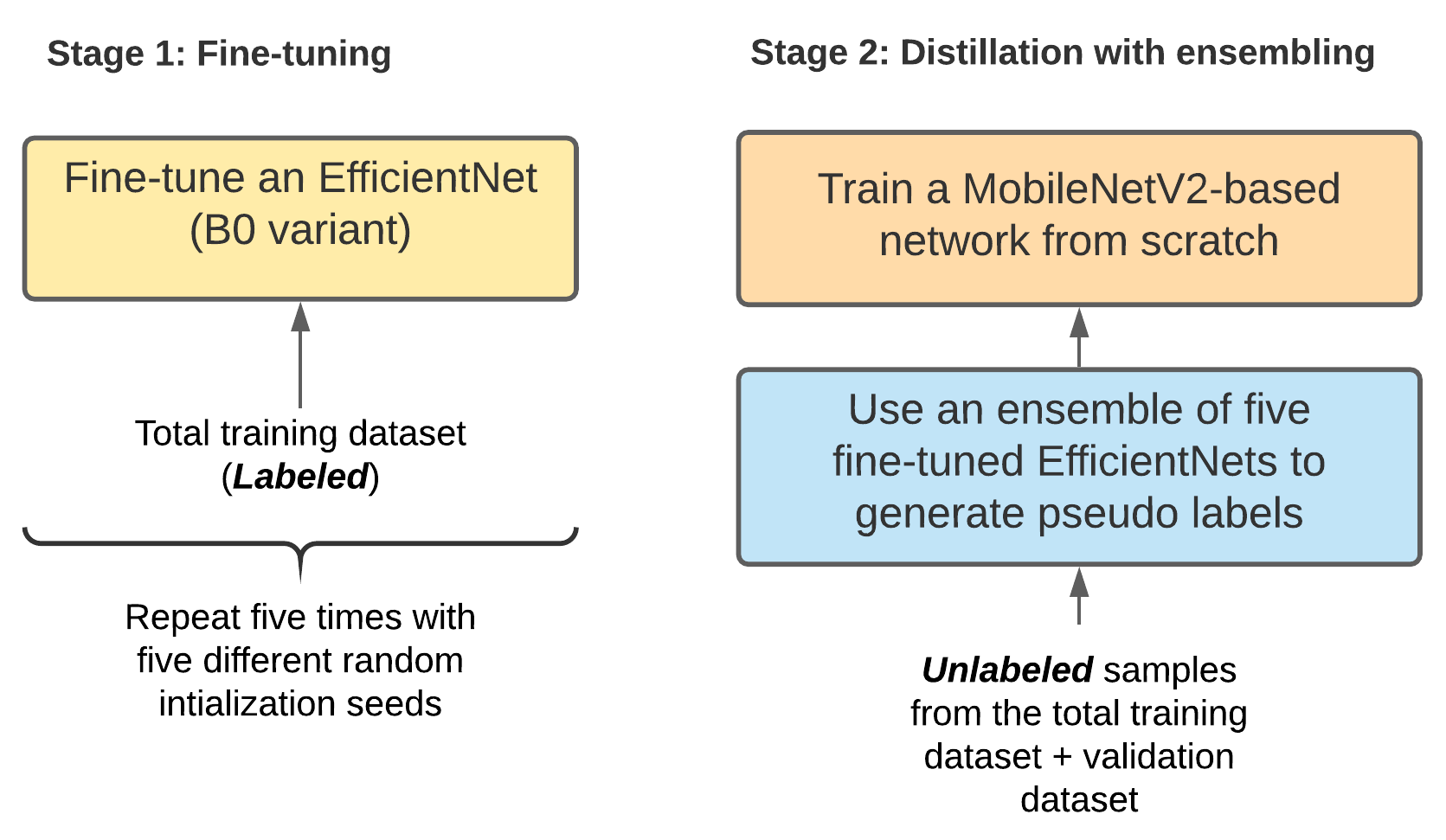}
}
\caption{\small{The solution proposed by PyImageSearch team.}}
\label{fig:thanki_diagram}
\end{figure}

\begin{figure}[h!]
\centering
\resizebox{1.0\linewidth}{!}
{
\includegraphics[width=1.0\linewidth]{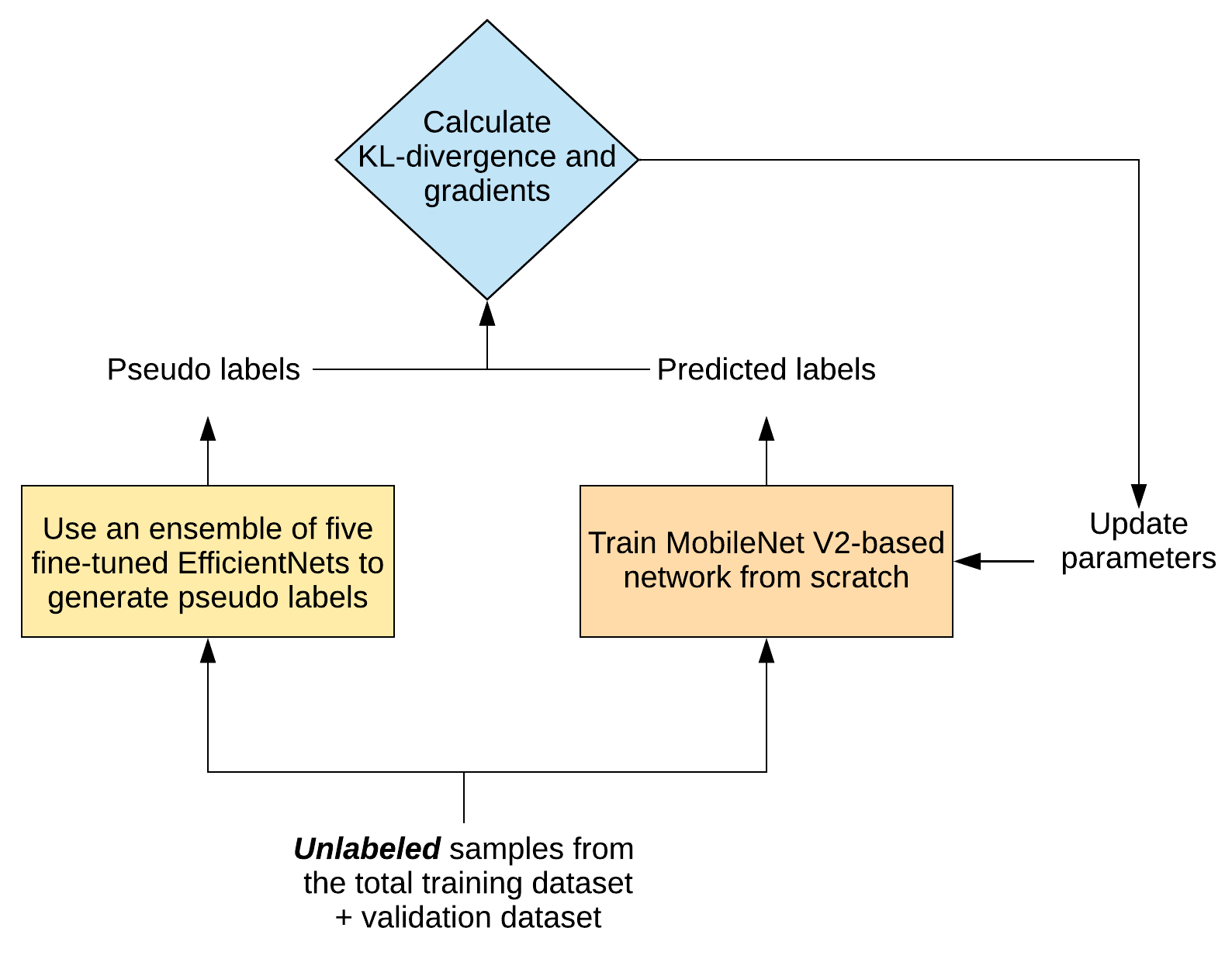}
}
\vspace{0.2mm}
\caption{\small{Knowledge distillation approach used by PyImageSearch.}}
\label{fig:thanki_distillation}
\end{figure}

Team PyImageSearch used the Noisy Student Training~\cite{xie2020self} approach (Fig.~\ref{fig:thanki_diagram}) to solve the considered classification problem. EfficientNet-B0 model~\cite{tan2019efficientnet} was initially fine-tuned on the training dataset using sparse categorical cross-entropy five times with different random initialization seeds. To train the target MobileNetV2-based model, an ensemble of five obtained EfficientNet-B0 networks was used as a source for pseudo labels. The target model was trained using the KL-divergence as the distillation loss function, the distillation process is illustrated in Fig.~\ref{fig:thanki_distillation}. Adam optimizer was used for training both teacher and student models. To mitigate the problem of class imbalance, the authors used class weights (calculated from the proportion of samples per class) when calculating the cross-entropy loss for training teacher models. Stochastic weight averaging (SWA) was additionally applied to utilize the previously updated parameters from earlier epochs, which induces an ensembling effect~\cite{izmailov2018averaging}. The student model was trained for 110 epochs with early stopping.

\subsection{Sidiki}

This is the only solution that is not using any backbone networks. The proposed model can achieve an accuracy of over 80\% on the validation set with only 60K parameters and the total model size of 73 KB. The model uses only 8 convolutional layers, 4 pooling layers and a fully connected layer. To speed up the inference, downsampling is applied in the first two convolutional layers by using strides 4 and 2 together with a small (6 and 12) number of channels.

The model was trained using Adam optimizer with a batch size of 32 as follows: for 10 epochs with a learning rate of $1e-3$, for 40 epochs with a learning rate of $1e-4$, then for 100 epochs with a learning rate of $7e-5$ and another 100 epochs with a learning rate of $9e-5$. After that, all layers of the model were prepared for Quantization Aware Training and then the model was trained for another 100 epochs with a learning rate of $1e-4$, and for 100  epochs~--- with $1e-5$. Once the final accuracy of 80\% was established on the validation set, the validation set was used for training of the quantized aware model for 50 epochs with a learning rate of $1e-4$. Then, the model was trained again on the original training set for 10 epochs with a learning rate of $1e-4$, and for 10~---  with $1e-6$. Finally, the model was repeatedly trained around 6-8 times on the training and validation sets for 10 and 1 epoch, respectively.

\section{Additional Literature}

An overview of the past challenges on mobile-related tasks together with the proposed solutions can be found in the following papers:

\begin{itemize}
\item Learned End-to-End ISP:\, \cite{ignatov2019aim,ignatov2020aim}
\item Perceptual Image Enhancement:\, \cite{ignatov2018pirm,ignatov2019ntire}
\item Image Super-Resolution:\, \cite{ignatov2018pirm,lugmayr2020ntire,cai2019ntire,timofte2018ntire}
\item Bokeh Effect Rendering:\, \cite{ignatov2019aimBokeh,ignatov2020aimBokeh}
\item Image Denoising:\, \cite{abdelhamed2020ntire,abdelhamed2019ntire}
\end{itemize}

\section*{Acknowledgements}

We thank AI Witchlabs and ETH Zurich (Computer Vision Lab), the organizers and sponsors of this Mobile AI 2021 challenge.

\appendix
\section{Teams and Affiliations}
\label{sec:apd:team}

\bigskip

\subsection*{Mobile AI 2021 Team}
\noindent\textit{\textbf{Title: }}\\ Mobile AI 2021 Challenge on Fast and Accurate Quantized Camera Scene Detection on Smartphones\\
\noindent\textit{\textbf{Members:}}\\ Andrey Ignatov$^{1,2}$ \textit{(andrey@vision.ee.ethz.ch)}, Grigory Malivenko \textit{(grigory.malivenko@gmail.com)}, Radu Timofte$^{1,2}$ \textit{(radu.timofte@vision.ee.ethz.ch)}\\
\noindent\textit{\textbf{Affiliations: }}\\
$^1$ Computer Vision Lab, ETH Zurich, Switzerland\\
$^2$ AI Witchlabs, Switzerland\\

\subsection*{ByteScene}
\noindent\textit{\textbf{Title:}}\\ Transfer Knowledge from Both Big Pre-trained Models and Small Pre-trained Models\\
\noindent\textit{\textbf{Members: }}\\ \textit{Sheng Chen (chensheng.lab@bytedance.com)}, Xin Xia, Zhaoyan Liu, Yuwei Zhang, Feng Zhu, Jiashi Li, Xuefeng Xiao, Yuan Tian, Xinglong Wu \\
\noindent\textit{\textbf{Affiliations: }}\\
ByteDance Inc., China \\

\subsection*{EVAI (Edge Vision and Visual AI)}
\noindent\textit{\textbf{Title:}}\\ Scene Detection with truncated MobileNetV2 Backbone \\
\noindent\textit{\textbf{Members: }}\\ \textit{Christos Kyrkou (ckyrko01@ucy.ac.cy)}\\
\noindent\textit{\textbf{Affiliations: }}\\
KIOS Research and Innovation Center of Excellence, University of Cyprus, Cyprus \\

\subsection*{ALONG}
\noindent\textit{\textbf{Title:}}\\ EfficientNet-Lite4 for Real-Time Scene Detection \\
\noindent\textit{\textbf{Members: }}\\ \textit{Yixin Chen (chenyixin03@corp.netease.com)}, Zexin Zhang, Yunbo Peng, Yue Lin \\
\noindent\textit{\textbf{Affiliations: }}\\
Netease Games AI Lab, China \\

\subsection*{Team Horizon}
\noindent\textit{\textbf{Title:}}\\ MobileNetV2 with Resized Input for Fast Scene Detection \\
\noindent\textit{\textbf{Members: }}\\ \textit{Saikat Dutta$^1$ (saikat.dutta779@gmail.com)}, Sourya Dipta Das$^2$, Nisarg A. Shah$^3$, Himanshu Kumar$^3$ \\
\noindent\textit{\textbf{Affiliations: }}\\
$^1$ Indian Institute of Technology Madras, India\\
$^2$ Jadavpur University, India\\
$^3$ Indian Institute of Technology Jodhpur, India \\

\subsection*{Airia-Det}
\noindent\textit{\textbf{Title:}}\\ Bag of Tricks for Real-Time Scene Detection Based on MobileNets \\
\noindent\textit{\textbf{Members: }}\\ \textit{Chao Ge (stvea@qq.com)}, Pei-Lin Wu
Jin-Hua Du \\
\noindent\textit{\textbf{Affiliations: }}\\
Nanjing Artificial Intelligence Chip Research, Institute of Automation, Chinese Academy of Sciences, China \\

\subsection*{DataArt Perceptrons}
\noindent\textit{\textbf{Title:}}\\ Using MobileNetv2 for scene detection \\
\noindent\textit{\textbf{Members: }}\\ \textit{Andrew Batutin (andrey.batutin@dataart.com)}, Juan Pablo Federico, Konrad Lyda, Levon Khojoyan \\
\noindent\textit{\textbf{Affiliations: }}\\
DataArt Inc, United States \\

\subsection*{PyImageSearch}
\noindent\textit{\textbf{Title:}}\\ A Bag of Tricks for Mobile-Friendly Image Classification \\
\noindent\textit{\textbf{Members: }}\\ \textit{Sayak Paul (s.paul@pyimagesearch.com)}, Abhishek Thanki \\
\noindent\textit{\textbf{Affiliations: }}\\
PyImageSearch, India \\

\subsection*{Sidiki}
\noindent\textit{\textbf{Title:}}\\ Tiny Mobile AI Model \\
\noindent\textit{\textbf{Members:}}\\ \textit{Shahid Siddiqui (msiddi01@ucy.ac.cy)} \\
\noindent\textit{\textbf{Affiliations: }}\\
KIOS Center of Excellence, University of Cyprus, Cyprus\\

{\small
\bibliographystyle{ieee_fullname}

}

\end{document}